# Hardware Software Co-design of the Aho-Corasick Algorithm: Scalable for Protein Identification?

S.M. Vidanagamachchi, S.D. Dewasurendra, *Member, IEEE* and R.G. Ragel, *Member, IEEE*

*Abstract*-Pattern matching is commonly required in many application areas and bioinformatics is a major area of interestthat requires both exact and approximate pattern matching.Much work has been done in this area, yet there is still a significant space for improvement in efficiency, flexibility, and throughput.This paper presents a hardware-software co-design of Aho-Corasick algorithm in NiosII soft processor and its scalability for a pattern matching application. A software only approach has used to compare the scalability of hardware-software co-design and according to the results we could conclude that software-hardware co-design implementation could give a maximum of 10 times speed up for automata created with 1200 peptides compared to software-only implementation.

## I. INTRODUCTION

Due to the fast growth of biological sequence data in biological databases, pattern matching in bioinformatics demands exceptionally high performance. Over the past few decades a lot of effort has gone into filling the gap between data generation speed and the data processing speed.One particularly effective approach in this context has been based oncomputational resources such as soft-core processors which provide flexible platforms for computationally complex problems through the reconfigurable nature of Field Programmable Gate Arrays (FPGAs).

Commercially available Altera and Xilinx FPGAs provide reprogrammablesoft-core processors named Nios II and MicroBlaze, respectively. We could use these FPGAplatforms for developing algorithms that could be further optimized with soft-core processors (Nios IIin our case).It is envisioned that reconfigurable processors will play an important role in future embedded SoC platforms due to their ability to deal with the technological and market challenges. A reprogrammable processor can support implementation of critical parts of an application in hardware using a specialized instruction set. Here we have developed a hardware-software co-design using System On a Programmable Chip (SOPC) component implementation with Altera provided Nios II processorto solve a common exact string matching problem in Computational Biology.

The objectives of this paper are to describe hardware-software co-design of Aho-Corasick algorithm with Nios II Embedded processor and to demonstratethe performance enhancement and scalability achieved through thiscustom implementation ofAho-Corasickalgorithm over a software-only implementation.

Aho-Corasickalgorithm is a widely used multiple string matching algorithm and it has a linear time complexity: O(sum of the lengths of keywords) in the pre-processing stage and O(Total length of the text string to be processed) in the matching stage[10]. We have made two implementations ofthe Aho-Corasick algorithm: one, a software only implementation on a Nios II processor with its usual instruction set and the other a hardware software co-design implementation with extended instruction set for Nios II processor.That is, thecomputationally intensive part of the algorithm is incorporated into the soft-coreand run through an instruction,whichmakes the latter a hardwaresoftware co-design. We have then measured the speed up we achieved with the hardware-software co-design compared to the software only implementation.

Hardware-software co-design refers to the development of heterogeneous systems, where computationally expensive parts (such aslarge matrix multiplication, FSM logic, etc.) could be implemented as custom components/hardware. It takes the advantage of both the flexibility of the soft-core processor and the speed and the power of dedicated hardware.

In our experiments, both of the hardware software co-design and the software onlyimplementations are aimed at matching peptides in a selected set of long protein sequences. Fast peptide matching is useful in several biological processes such as identifying disease causing peptides/proteins in any organism, finding homologous/evolutionary patterns in organisms and a host of other useful studies [11][12].

The rest of this paper is organized as follows.InSection II we have included the literature of softwarehardware co-design of algorithms, speciallyrelated to string matching.Methodology followed inour experiments isdescribed in Section III. In Section IV we have shown our results andconclusionis presented in Section V.

## II. LITERATURE REVIEW

Hardware software co-designs for several types of algorithms have been reported in the literature:an implementation for a line detection algorithm hasbeendeveloped by Kayankit et al. and was simulated in [1], a face recognition algorithm was implemented by Ming et al. [2]and a QR Decomposition based Recursive Least Square algorithm (an adaptive filter algorithm) was implemented by Nupur et al.[3]. A few implementations have been developed for string matching too using hardwaresoftware co-design principles:Hashmi et al. have used hardwaresoftware co-design for a snort detection system wherein they have useda

Bloom Filter to compact the large number of patterns needing storage[4]; ahash algorithm was implemented in[5] to locate millions of 100 base-pairs in a 3 million base-pair reference genome; a hardwaresoftware co-design was implemented by Ying-Dar et al. for signature based virus scanning in the ClamAV antivirus package which uses the Wu-Manber and Aho-Corasick algorithms [6];a new, platform independentapproach to FSM (Finite State Machine) based KMP (Knuth Morris Pratt) algorithm hasbeen introduced by Nader et al. in[7] and a mapping of the EffiCuts algorithm to the PLUG platform was performed by Vaish et al. andthey reportedlyhave achieved high throughput and low power [8].Table I shows the differencesbetween some of theimplementations discussed here and our implementation of apeptide matchingalgorithm.

TABLE I
COMPARISON OF RELATED WORK

|  | **Snort detection engine [4]** | **Signature based virus scanning [6]** | **Packet classification [8]** | **Pattern matching hardware [7]** | **Peptide identification (Our system)** |
|---|---|---|---|---|---|
| **Hardware** | Accelerate malware pattern search, use Bloom filter based accelerator hardware module in Xilinx Virtex II FPGA operating at 50MHz | Accelerate signature matching, use BFAST (Bloom Filter Accelerated Sublinear Time) architecture in Xilinx ML310 Virtex II Pro Board operating at 100MHz | Accelerate packet classification using EffiCuts algorithm which has map to PLUG hardware architecture in Xilinx Virtex-5 XC5VFX200T at 550 MHz | Accelerate string matching with Knuth Morris Pratt algorithm in Xilinx Spartan 3E Starter board at 50 MHz | Accelerate peptide matching, Use Aho-Corasick algorithm in Altera DE2 FPGA operating at 50MHz |
| **Software** | Rule parsing, rule selection and packet header checking functions, run in Micro Blaze soft processor with HandelC operating at 50MHz | Verification module is implemented using Wu Manber and Aho-Corasick algorithms (ClamAV), run Xilinx ML310 Virtex II Pro Board soft CPU in operating at 300MHz | Task scheduling to access network or memory banks Micro Blaze processor at 550 MHz | FSM construction and reconfiguration in Micro Blaze soft processor at 50 MHz | Handling input output functions in Nios II soft processor operating at 50MHz |
| **Result** | Have best effective throughput of 2.49Gbps and lowest throughput of 1.71Gbps | Have throughput of 151 Mbps (28.1 times faster than original ClamAV's throughput) | For packet sizes of 64 bytes, throughput is 71 Gbps | Have approximate speed up of 7 times compared to multi context FPGA | Obtained a speed up of 10 times compared to software only implementationin Nios II processor |

III. METHODOLOGY

Aho-Corasick is arguablythe best and the widest used multiple pattern matchingalgorithmthat searches all occurrences of any of a finite number of keywords in a text string. Dandass et al. have used this algorithm for hardware acceleration of peptide pattern matching for the first chromosome of human genome [9]. We have used this algorithm to optimizing the areausage of FPGA for peptide matching [14].This algorithm consists of two phases; constructing a finite state machine from keywords and then using these state machines for locating the keywords by processing the text string in a single pass. In their hardware based implementation Dandass et al. have mapped 2800 peptides into a Xilinx FPGA and compared it with a workstation implementation. In our custom component based hardware software co-design implementation, we have usedthe maximum number of peptides which can be identified with our FPGA. Our custom componentconsists of a hardware implementation of line 3 in the pseudo-code of Algorithm 1:i.e., `Do-Matching-Process()`,which corresponds tothe matching phase of Aho-Corasick algorithm.TheAho-Corasick Finite State Machines aregenerated in the pre-processing stagebased on thepeptide patternsto be considered. Our custom component design with Avalon memory mapped interface is as follows (Figure 1).

Avalon Memory Mapped interface is a bus like protocol which facilitates the communication between the Nios II processor and the user defined custom component.It provides address based read/write interfaces to support off-chip peripherals.When byte enable signal of Avalon Memory Mapped Slave interface is high,memory mapped slave selects one of its registers to read (store) input characters which arewritten to itthrough a software macro in Nios II Integrated Development Environment (IDE)and subsequently, to write back the results of matching process to thesame register. Effectively, we send the protein list(concatenating several proteins)in and get the result back via Nios II IDE.

Quartus II software provides theSOPC builder feature which facilitatesa number of components to be integrated tomake a complete system. Avalon Switch Fabric (an interconnection network) is the interface to connect all these peripherals and custom peripherals/components. Thus, Quartus II software and Nios II Embedded Design Suit (EDS) could be used together to develop hardwaresoftware co-designs of specific applications. In our co-design, thehardware component accelerates the function of matching peptides and the software component provides the input and output interface for the system.

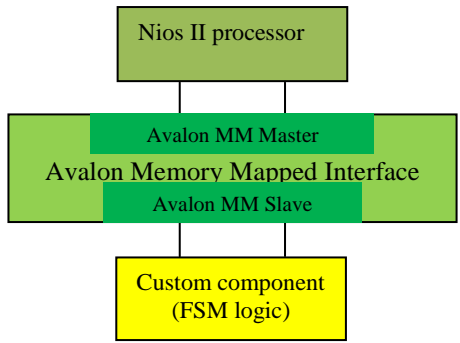

Fig.1. Add custom component to Nios II Processor

We have mapped Aho-Corasick finite state machines in hardware, for which we have used a tool presented in [13] and the stages of the hardware design automation are shown in Figure 2. When we input the extracted protein datasets in software, it uses the FSM designed in hardware to search through and locate eachpeptide occurrence in the proteins. This software acts as an interface to send inputs via Avalon Memory Mapped(AMM) interface and finally gets back the result via the same AMM interface. In the software, the invocation of hardware component for string matching is performed by macro instructions and these instructions are run iteratively to process a given input protein (once per character).Algorithm 1 is the pseudo-code of the whole process.

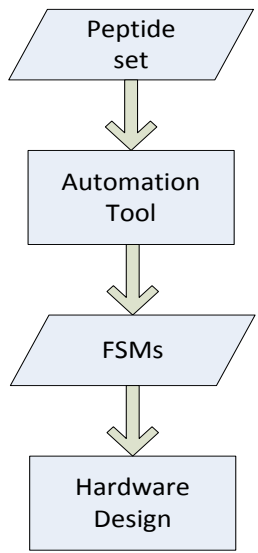

Fig.2. Stages of hardware design automation

As the first step in the hardwaresoftware co-design implementation of the Algorithm 1 we write the input string (corresponding to the protein to be identified) character by character to the Memory Mapped Slave register until end of the string (lines 1 and 2). Then the hardware performs the matching process (line 3). Finally the result is written back to the Memory Mapped Slave register (line 4) which can be read by the software for analysis.

Then the same algorithm was implemented only in software (using Nios II instruction set) and we measured the corresponding time for each protein set and peptide set while increasing the numbers in both sets. Software only system was implemented and runs in the Nios II soft-core processor. This was then comparedwith the hardwaresoftware co-design implementation.

```
1. While(Not(EndOfProteinList)){
2.    WriteA-Character-To-MM-Slave-Register()
3.    Do-Matching-Process()
4.    Write-Result-To-MM-Slave-Register()
5. }
```

Algorithm 1. Algorithm for hardware-software co-design process

### IV. DATA COLLECTION AND PRE-PROCESSING

We have used protein data from UniProt database and PeptideMass software which are available in Expasy Proteomics Server[1]. PeptideMass creates possible peptides for a given protein. In the pre-processing step of the hardwaresoftware co-design these peptides areinput intoa software tool (which is implemented in C++) to generate Finite State Machines (FSMs) inVHDL (Very-high-speed integrated circuit Hardware Description Language):wepresented this tool in [13].

TABLE II
NUMBER OF AMINO ACIDS IN EACH PROTEIN SET

| Number of proteins | 100 | 500 | 1000 |
|---|---|---|---|
| Number of characters | 53093 | 172141 | 329527 |

We have matched the peptide set against 100, 500 and 1000 protein data setsrespectively and recorded the times taken for execution. Table II lists the number of amino acids available in the protein data sets we have considered. For instance, the total number of amino acids in the protein data set of 500 proteins was 172141.

### V. RESULTS AND DISCUSSION

Table III and IV show the matching times (in μs) for hardwaresoftware co-design and software only implementations for the complete matching operation respectively.

---

[1] http://web.expasy.org/peptide_mass/

TABLE III
TIME TAKEN (in μs) FOR MATCHING IN HARDWARE SOFTWARE CO-DESIGN IMPLEMENTATION

| Pep. / Pro. | 100 | 500 | 1000 | 1200 |
|---|---|---|---|---|
| 100 | 59209 | 59209 | 59171 | 59167 |
| 500 | 192637 | 192646 | 192602 | 192602 |
| 1000 | 367323 | 367323 | 367323 | 367137 |

Experiments were performed with protein sets of sizes 100, 500 and 1000. We have used 100, 500, 1000 and 1200 peptide sets for the automata (FSM) creation.

According to the time taken for the matching process of hardwaresoftware co-design (Table III), for a fixed length of protein set, different automata which are developed using 100, 500, 1000 and 1200 peptides show approximately equal time for the matching process. For example, if we take protein data set 100, it takes approximately 59200 μs to match over the automata of 100, 500, 1000 and 1200 peptides. In hardware software co-design, we have two fixed macro instructions to be executed in order to process each character/amino acid in the protein. Since the input length of the protein is fixed for a particular row in Table III, to match over different sizes of automata, we have to perform same number of instructions per each automaton independent of the size of the automaton. Therefore, for each automaton on a row, the same numbers of clock cycles are used to perform the matching process. Therefore we get approximately equal matching time in each row in Table III.

TABLE IV
TIME TAKEN (IN μs) FOR MATCHING IN SOFTWARE ONLY IMPLEMENTATION

| Pep. / Pro. | 100 | 500 | 1000 | 1200 |
|---|---|---|---|---|
| 100 | 311 424 | 357 435 | 446 742 | 625 682 |
| 500 | 1 018544 | 1 164 763 | 1 616 889 | 1 848 193 |
| 1000 | 1 931 712 | 2 231 601 | 3 102 777 | 3 428 774 |

In Table III, if we consider column wise the value is increasing when we increase the number of proteins. When we increase the number of proteins, the number of amino acids to match will also increase as it was shown in Table II. Therefore, the time taken will increase and the time taken for 500 proteins is approximately 3 times compared to time taken for 100 proteins. It is because the length of the 500 proteins is 3 times compared to 100 proteins (the length is calculated based on the total number of amino acids as shown in Table II). Further, the time taken for 1000 proteins is approximately 2 times compared to time taken for 500 proteins due to the same reason. In fact this behaviour is consistent with the expression for running time of Aho-Corasick algorithm; viz. O (length of the text string to be processed).

According to the results in Table IV of software only implementation, the results of each row from left to right increase with the number of peptides. That could bebecausethe individual AC trees with failure links (which then becomes a labelled graph) are implemented as array of edge structures (a struct edge consists of the attributes: outgoing character and the next node) (Figure 3). When the number of peptides included increases, the number of elements in the array at each node of the graph tends to increase considerably and hence, the time taken to search through each of these elements (for the next input symbol in the input string) also increases. This array can also be implemented as an adjacency list. In fact the process of matching an input string against an AC graph is similar to reading the input string as an input tape to the graph considered as an automaton. Hence, when a protein sequence/s with a fixedlength is matched against (read into) automata of increasing sizes (100, 500, 1000 and 1200), one can expect the search time at each node to increase. The increase in processing times coming down a column in the table is easily explained by the increased length of input string (running time of Aho-Corasick algorithm = O (Total length of the text string to be processed)). The considerably increased processing times recorded for each case (e.g., the values 53093 and 311424, respectively, for the first cell in tables III and IV) indicates that the matching process in Hardware Software implementation, which is equal to the cycle time, is much shorter than the execution time of the corresponding set of instructions in the software only implementation.

TABLE V
SPEED UP FOR HARDWARE SOFTWARE CO-DESIGN TO SOFTWARE ONLY

| Pep / Pro | 100 | 500 | 1000 | 1200 |
|---|---|---|---|---|
| 100 | 5.26 | 6.04 | 7.55 | 10.57 |
| 500 | 5.29 | 6.05 | 8.39 | 9.60 |
| 1000 | 5.26 | 6.08 | 8.45 | 9.34 |

The speedup between the hardwaresoftware co-design and the software only implementation are presented in Table V. The speedup increases with increasing number of peptides, hence, with increasing size of Aho-Corasick finite state machines. This is due to the fact that the hardware software co-design scales with the increasing number of peptides while the software only approach does not scale. That is, the matching time in a hardware software co-design depends onlyon the length of the protein and not the size of the finite state machine (which is defined by the number of peptides in a peptide identification process).

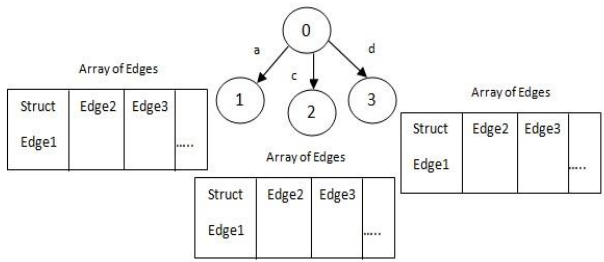

Fig.3. Data structure for keeping next state edges/nodes

According to the results shown in Table V, we could see that when we increase the number of peptides in automata from 100 to 1200 (100,500,1000 and 1200), in each protein set of sizes 100, 500 and 1000, respectively, we could achieve approximately 5, 6, 8 and 10 times speed up in hardwaresoftware co-design with respect to software only implementation.

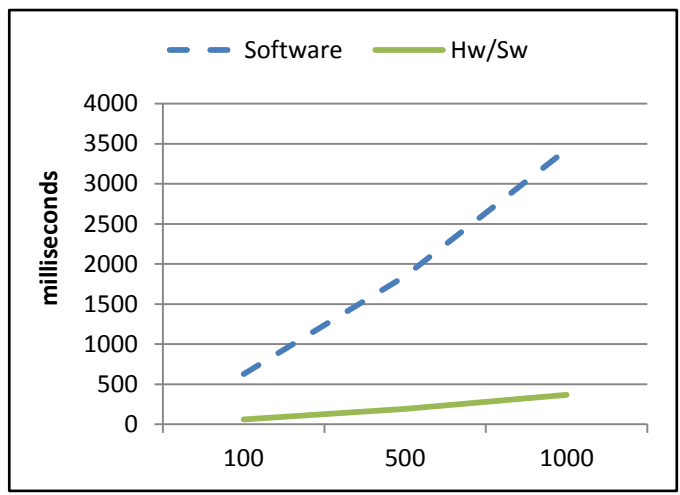

Fig.4. Total time for matching and speed up for 1200 peptides

In addition, in Figure 4, we have plotted the time taken for protein identification on both hardware software co-design (*Hw/Sw*) and software only (*Software*) approaches for a fixed FSM with changing input protein data sets. The FSM used for this plot is the 1200 peptide set FSM. As it was explained earlier, the time taken for protein identification in both hardware software co-design approach and the software only approach will increase with increasing number of proteins.Linearity can be seen in the plot of software only implementation clearly where time taken to process one character is exactly 1.1 micro seconds (due to exact number of instructions used for processing each character). In hardware software co-design it is changing from 10.4 to 11.7 micro seconds (due to change of number of instructions that are used to process each character).However, as it can be seen in Figure 3, the rate of increase for the hardware software co-design approach is much less compared to the rate of increase for the software only approach. Therefore, it could be concluded that the hardware software approach is much more scalable compared to the software only approach.

## VI. CONCLUSION

Protein identification is an area where improving the performance of string matching has great significance. In this paper, we have improved the efficiency of Aho-Corasick algorithm by using a hardware software co-design approach. In addition, in this paper we compared a hardwaresoftware co-design implementation with a software only implementation of Aho-Corasick algorithm in Nios II soft processor. According to the results we could conclude that hardwaresoftware co-design implementation could obtain a maximum of 10 times speed up for automata created with 1200 peptides compared to software-only implementation. In addition, we have also shown that the hardware software co-design approach is more scalable compared to the software only approach.